\documentclass[aps,pra,twocolumn,superscriptaddress]{revtex4}
\usepackage[caption=false]{subfig}
\usepackage{amssymb}
\usepackage{amsmath}
\usepackage{commath}
\usepackage{graphicx,bm}
\usepackage{verbatim}
\usepackage[usenames]{color}

\begin{document}
\title{Theory for Bose-Einstein condensation of light in nano-fabricated semiconductor microcavities}

\author{A.-W. de Leeuw}
\email{A.deLeeuw1@uu.nl}
\affiliation{Institute for Theoretical Physics and Center for Extreme Matter and Emergent Phenomena, Utrecht
University, Leuvenlaan 4, 3584 CE Utrecht, The Netherlands}
\author{E.C.I. van der Wurff}
\affiliation{Institute for Theoretical Physics and Center for Extreme Matter and Emergent Phenomena, Utrecht
University, Leuvenlaan 4, 3584 CE Utrecht, The Netherlands}
\author{R.A. Duine}
\affiliation{Institute for Theoretical Physics and Center for Extreme Matter and Emergent Phenomena, Utrecht
University, Leuvenlaan 4, 3584 CE Utrecht, The Netherlands}
\author{D. van Oosten}
\affiliation{Debye Institute for Nanomaterials Science and Center for Extreme Matter and Emergent Phenomena, Utrecht
University, Princetonplein 5, 3584 CC Utrecht, The Netherlands}
\author{H.T.C. Stoof}
\affiliation{Institute for Theoretical Physics and Center for Extreme Matter and Emergent Phenomena, Utrecht
University, Leuvenlaan 4, 3584 CE Utrecht, The Netherlands}
\date{\today}

\begin{abstract}
We construct a theory for Bose-Einstein condensation of light in nano-fabricated semiconductor microcavities. We model the semiconductor by one conduction and one valence band which consist of electrons and holes that interact via a Coulomb interaction. Moreover, we incorporate screening effects by using a contact interaction with the scattering length for a Yukawa potential and describe in this manner the crossover from exciton gas to electron-hole plasma as we increase the excitation level of the semiconductor. We then show that the dynamics of the light in the microcavities is damped due to the coupling to the semiconductor. Furthermore, we demonstrate that on the electron-hole plasma side of the crossover, which is relevant for the Bose-Einstein condensation of light, this damping can be described by a single dimensionless damping parameter that depends on the external pumping. Hereafter, we propose to probe the superfluidity of light in these nano-fabricated semiconductor microcavities by making use of the differences in the response in the normal or superfluid phase to a sudden rotation of the trap. In particular, we determine frequencies and damping of the scissors modes that are excited in this manner. Moreover, we show that a distinct signature of the dynamical Casimir effect can be observed in the density-density correlations of the excited light fluid. 
\end{abstract}

\pacs{67.85.Hj, 42.50.Ar, 47.37.+q}


\maketitle
\section{Introduction} 
The first observation of Bose-Einstein condensation in a dilute atomic vapor opened up several different possibilities to explore many-body phenomena in a completely new regime. As before this observation it was hard to experimentally access the macroscopic quantum regime, the low temperatures of these systems created a playground for the investigation of several interesting quantum effects. One prominent example is the observation of superfluidity via the existence of quantized vortices \cite{vort, vort1, vort2}. These developments were even more encouraged by the large experimental control that is available in the cold atomic gases, leading also to the possibility to investigate dynamical behavior.
\newline
\indent More recently a class of Bose-Einstein condensates have been observed that are fundamentally different from atomic condensates and also allow for different experimental probes. These Bose-Einstein condensates of bosonic quasiparticles such as magnons \cite{BECmagnon}, exciton-polaritons \cite{BECpolariton,BECpolariton2} and photons \cite{BECphoton} are dissipative systems. This leads to the possibility to investigate phenomena that are not not yet observed in atomic condensates, such as large number fluctuations in a Bose-Einstein condensate of photons \cite{Weitz2, NumFluc}. Moreover, it is also interesting to investigate whether certain equilibrium phenomena are still present in these dissipative systems.
\newline
\indent Although the Bose-Einstein condensate of light allows for dynamical measurements and new experimental probes, there are still some disadvantages to the current experimental approach \cite{BECphoton,Nyman}. First, it is inconvenient to dynamically change the trap geometry of the photons. Second, every possible change highly affects the interactions of the photons with the dye and therefore strongly influences the thermalization of the photons. Lastly, it is very hard to predict theoretically, and to a large extent it is still unknown experimentally, how strongly the photons interact and what the origin of the effective photon-photon interaction is. Therefore, this is a large disadvantage for studying various many-body phenomena of light.
\newline
\indent To overcome these problems we propose to follow a different experimental approach, namely nano-fabricated semiconductor microcavities. In this system we use a semiconductor with a periodic array of holes filled with air to create a photonic crystal. The photons thermalize due to the interaction with the electrons and holes in the semiconductor. Moreover, by simply increasing the size of the holes when moving further away from the center of the semiconductor, the photons feel an effective harmonic trapping potential \cite{Sebas}. Finally, also in this case there is external pumping to compensate for photon losses out of the microcavity. Although this suggest that light in such nano-fabricated semiconductor microcavities is similar to photons in a dye-filled optical microcavity, the former system has several advantages. Most importantly, the interaction between the light and the semiconductor is well understood, see e.g. Ref.\,\cite{BookSC}. Therefore, there are good prospects of controlling the effective photon-photon interaction. Moreover, it is possible to change the cavity geometry and it is also achievable to influence the photon trap dynamically on a sub-picosecond timescale \cite{Harding}.
\newline
\indent 
\indent A prime example of a many-body phenomenon of light where the advantages of the semiconductor microcavity can be utilized is superfluidity. In Ref.\,\cite{Keeling} it is shown that in dissipative condensates a superfluid density can exist and therefore it is expected that also a Bose-Einstein condensate of photons can exhibit superfluidity. Although some work has been carried out that mainly focus on the frictionless flow of light through an obstacle, see e.g. Ref.\,\cite{sup1,sup2,sup3,sup4}, up to now the superfluidity of a Bose-Einstein condensate of photons has not yet been explored, since it is unclear how the superfluidity of this system can be probed in the current experiments. Namely, recall that the standard experiment for probing the superfluid behavior, i.e., observing the existence of quantized vortices after rotation of the condensate, appears not to be feasible in a dye-filled optical microcavity.
\newline
\indent From studies in atomic Bose-Einstein condensates we know that there is another method to obtain direct evidence for superfluidity \cite{SM1,SM2}. By applying a small sudden rotation of the trapping potential the so-called scissors modes can be excited. The modes are quite different from the excitations in the of the atoms in the normal state. Therefore, the time evolution of the angle between the axial direction of the condensate and the new trap direction is distinct in the normal and superfluid phase. As mentioned before, in current experiments this is difficult to observe as a rotation of the cavity highly affects the interactions of the photons with the dye and therefore destroys the thermalization of the photons. In contrast, in a semiconductor microcavity the trap geometry can be changed on a sub-picosecond timescale \cite{Harding}, which is much shorter than the inverse trapping frequency which is typically on the order of tens of gigahertz. Therefore, this fast change of the cavity geometry is non-adiabatic, which allows for the excitation of the scissors modes and probing the superfluidity of a Bose-Einstein condensate of light. 
\newline
\indent Nonetheless, there are still major differences between the dynamics of the scissors modes in Bose-Einstein condensates of atoms and photons. The former systems are very clean and therefore damping is typically not an issue, especially because most experiments are carried out in the collisionless regime. As a result, only for fine-tuned configurations of the experimental set-up, such as in Ref.\,\cite{Bel}, the scissors modes are damped by Beliaev processes. In a Bose-Einstein condensate of light on the other hand, the photons are, as envisaged here, coupled to electrons and holes in the solid-state cavity, that is pumped with an external laser beam. In the context of exciton-polariton systems it is shown that this pumping of the external bath affects the superfluid properties \cite{RefEP1, RefEP2}. We show here that the coupling with the external bath gives rise to damping of the scissors modes and in addition that this damping offers the possibility to observe a dynamical Casimir effect. Moreover, we demonstrate that the amount of damping also depends on the external pumping. 
\newline
\indent In this article we study light in nano-fabricated semiconductor microcavities. The layout of the paper is as follows. In Sec.\,\ref{sec:Semiconductor} we model the semiconductor by one conduction and one valence band and describe the interactions between the electrons and holes by an effective contact interaction, which is especially appropriate in the electron-hole
plasma regime of interest to us here where the Coulomb potential is screened to a short-range interaction. We include screening effects by calculating the scattering length for the appropriate Yukawa potential. Hereafter, we consider the coupling between light and the semiconductor in Sec.\,\ref{sec:Light}. We show that the coupling results into damping of the light and we demonstrate that this damping can be characterized by a single dimensionless parameter that depends on the external pumping. After this, we apply our model and study an example of a many-body phenomena of light where the advantages of semiconductor microcavities are particularly useful. Namely, in Sec.\,\ref{sec:Scissors} we propose to investigate the superfluidity of light via the excitation of scissors modes. In Sec.\,\ref{sec:DampingScissors} we consider a sudden rotation of the trapping potential of an elongated photon condensate and we calculate the decay rates of the excitations in the Thomas-Fermi limit. Hereafter, in Sec.\,\ref{sec:DDcor} we propose to measure the density-density correlations of the excited light fluid, since the decay products of the scissors mode quanta also demonstrate an analog of the dynamical Casimir effect. Finally, we end with conclusions and discussion in Sec.\,\ref{sec:concl}.

\section{Semiconductor} 
\label{sec:Semiconductor}
To investigate Bose-Einstein condensation of light in a nano-fabricated semiconductor we in first instance ignore the light and start with a model for a homogeneous semiconductor. We consider the following action
\begin{align}\label{eq:acsm}
S_{\mathrm{sc}}&[\phi^{*},\phi] \\ \nonumber
&= - \hbar \sum_{i,\alpha} \int_{0}^{\hbar\beta} d\tau \int d{\bf x} \, \phi^{*}_{i,\alpha}({\bf x},\tau) G^{-1}_{0i} \phi_{i,\alpha}({\bf x},\tau) \\ \nonumber
&- \sum_{\alpha, \alpha^{\prime}}  \int_{0}^{\hbar\beta} d\tau \int d{\bf x} \, d{\bf x}^{\prime}\, \phi^{*}_{e,\alpha}({\bf x},\tau) \phi^{*}_{h,\alpha^{\prime}}({\bf x}^{\prime},\tau) \\ \nonumber
&\times V_{s}({\bf x} - {\bf x}^{\prime}) \phi_{h,\alpha^{\prime}}({\bf x}^{\prime},\tau) \phi_{e,\alpha}({\bf x},\tau).
\end{align} 
In this model we only take into account one valence and conduction band and $i$ denotes the electron $e$ or hole $h$ respectively. The generalization to for instance one conduction band
and three valance bands is straightforward and can be easily achieved once the experimentally relevant semiconductor is known. The electron field and hole field are denoted by $\phi_{i,\alpha}$ and $\phi^{*}_{i,\alpha}$ with $\alpha$ and $\alpha^{\prime}$ representing the spin degeneracy that is denoted by $\uparrow$ or $\downarrow$. The noninteracting Green's function $G^{-1}_{0i}$ is defined as
\begin{align}
G^{-1}_{0i} = -\frac{1}{\hbar} \left\{\hbar \frac{\partial}{\partial \tau} - \frac{\hbar^{2} \nabla^{2}}{2 m_{i}} - \mu_{i} \right\},
\end{align}
where $\mu_{i}$ is the chemical potential and $m_{i}$ is the corresponding mass of the electron or hole. Note that the band gap of the semiconductor is absorped in our definitions of the chemical potentials. The same is true for the bandgap renormalization due to the electron-electron and hole-hole Coulomb interactions. The electron-hole interaction, however, needs to be
considered explicitly due to the possibility of exciton formation.
\newline
\indent In a semiconductor there are usually Coulomb interactions. However, since we are mostly interested in the highly excited regime where excitons do not exist and screening plays a dominant role, we may simplify the theory and replace the interaction potential by a contact interaction
\begin{align}
V_{s}({\bf x} - {\bf x}^{\prime}) \rightarrow -V_{0} \delta({\bf x} - {\bf x}^{\prime}),
\end{align}
with $V_{0}$ the effective interaction strength that we determine selfconsistently later on. 
\newline
\indent Now we perform a Hubbard-Stratonovich transformations to the pairing fields and integrate out the fermionic fields \cite{Marijn}. Note that we have to perform four different transformations and therefore we introduce the fields $\Delta_{\alpha \alpha^{\prime}}({\bf x}, \tau)$, of which the averages are given by
\begin{align}
\langle \Delta_{\alpha \alpha^{\prime}}({\bf x}, \tau) \rangle &= V_{0} \langle \phi_{h, \alpha} ({\bf x}, \tau)  \phi_{e,\alpha^{\prime}} ({\bf x}, \tau)  \rangle.
\end{align}
We apply here the Nozi\`{e}res-Schmitt-Rink approximation and only consider terms up to quadratic order in the pairing fields \cite{NSR}, which is the simplest approximation that correctly incorporates the crossover between an exciton gas and an electron-hole plasma that occurs as a function of excitation, i.e., pumping, of the semiconductor. In this approximation we find that the thermodynamic potential $\Omega$ is given by
\begin{align}
\Omega &:= \Omega_{1} + \Omega_{2} = \frac{4}{\beta} \sum_{{\bf P},n} \mathrm{ln}\left(- 1 / \hbar T^{\mathrm{MB}}(i \Omega_{n},{\bf P})\right) \\ \nonumber
&-\frac{2}{\beta} \sum_{{\bf p},i} \mathrm{ln}\left(1 + e^{-\beta (\epsilon_{{\bf p},i} - \mu_{i})}\right),
\end{align}
where $\Omega_{n}$ are bosonic Matsubara frequencies and $\epsilon_{{\bf p},i} = \hbar^{2} {\bf p}^{2} / 2 m_{i}$ is the kinetic energy of the particle or hole. The thermodynamic potential thus consists of the sum of the ideal electron and hole contributions and a fluctuation correction. The many-body T-matrix in the above expression is defined as
\begin{align}\label{eq:TM}
&\frac{1}{T^{\mathrm{MB}}(i \Omega_{n},{\bf P})} = \frac{1}{V_{0}}  - \frac{1}{V} \sum_{{\bf p}^{\prime}} \frac{1}{\epsilon_{{\bf p}^{\prime},e} + \epsilon_{{\bf p}^{\prime},h}} \\ \nonumber
&+ \frac{1}{V} \sum_{{\bf p}^{\prime}} \frac{1 - N_{\mathrm{FD}}(\epsilon_{{\bf P} - {\bf p}^{\prime},e} - \mu_{e}) - N_{\mathrm{FD}}(\epsilon_{{\bf p}^{\prime},h} - \mu_{h})}{-i \hbar \Omega_{n} + \epsilon_{{\bf P} - {\bf p}^{\prime},e} + \epsilon_{{\bf p}^{\prime},h} - \mu_{e} - \mu_{h}},
\end{align}
where $N_{\mathrm{FD}}(x) = 1 / (e^{\beta x} + 1)$ is the Fermi-Dirac distribution function. From now on we simplify the notation by omitting the arguments of the many-body T-matrix.

\subsection{Interactions} 
To obtain a better understanding of the effect of the interactions in our model, we now focus on the many-body T-matrix.
\begin{figure*}[t]
\centerline{\includegraphics[scale=1]{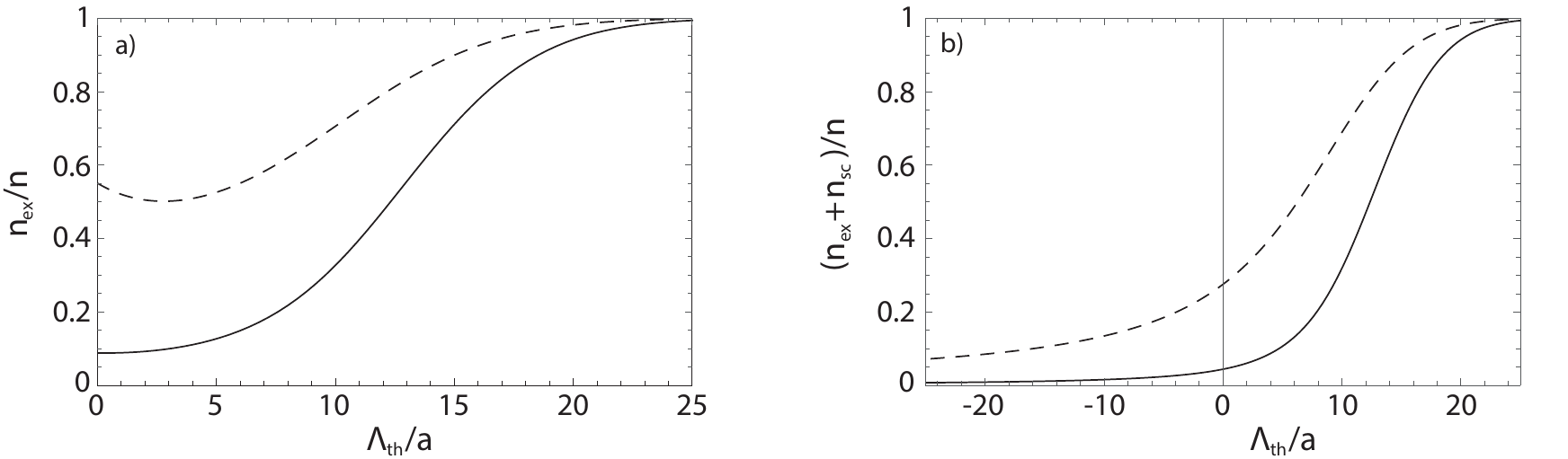}}
\caption{a) Number of electrons in excitons and b) total number of electrons in the excitons and the scattering contributions as a function of the inverse of the interaction parameter $a$ for ZnO. The solid and dashed line correspond to a carrier density of $n = 10^{23}\,\mathrm{m}^{-3}$ and $n = 10^{24}\,\mathrm{m}^{-3}$ respectively. We obtain a smooth crossover from $\Lambda_{\mathrm{th}} / a \ll 0$ where the electrons and holes behave as an ideal gas and form an electron-hole plasma, to $\Lambda_{\mathrm{th}}/ a \gg 0$ where almost all electrons and holes form excitons.}
\label{fig:BECBCS}
\end{figure*}
For simplicity we consider relatively small densities of electrons and holes, such that the many-body effects, i.e., the Fermi-Dirac distribution function in Eq.\,\eqref{eq:TM}, can be neglected. In this regime we find
\begin{align}\label{eq:2tm}
\frac{1}{T^{\mathrm{MB}}} = \frac{1}{V_{0}} - \frac{m_{r}^{3/2}}{2 \pi \hbar^{3}} \sqrt{-2 (z - \epsilon_{{\bf P},e + h} + \mu_{e} + \mu_{h})},
\end{align}
where $z$ is the complex center-of-mass energy associated with the Matsubara frequencies $i\Omega_{n}$ and $m_{r}$ is the so-called reduced mass which is defined as
\begin{align}
m_{r} = \left(\frac{m_{e} m_{h}}{m_{e} + m_{h}}\right).
\end{align}
Moreover, $\epsilon_{{\bf P},e + h}$ is the kinetic energy of the center-of-mass motion of the electron and hole involved in the interaction that is given by
\begin{align}\label{eq:encom}
\epsilon_{{\bf P},e + h} = \frac{\hbar^{2} {\bf P}^{2}}{2 (m_{e} + m_{h})}.
\end{align}
We write the effective interaction strength as $V_{0} = 4 \pi \hbar^{2} a / m_{r}$ with $a$ the so-called effective scattering length of the potential between the electron and hole. 
\newline
\indent From the expression for the many-body T-matrix in Eq.\,\eqref{eq:2tm} we obtain two features as a function of the complex energy $z$. First, we find that for positive values of $V_{0}$ there is a pole on the real axis. This corresponds to the exciton that is a bound state of an electron and a hole. Second, we find that there is a branch cut on the real axis, which corresponds to the continuum of electron-hole scattering states. To see this explicitly we calculate the density of electrons in the semiconductor. As we only consider optical excitations, the electron density $n_{e}$ is equal to the hole density $n_{h}$. Hence, we define a carrier density $n$ as
\begin{align}
n = n_{e} = n_{h} =  -\frac{1}{V} \left(\frac{\partial \Omega}{\partial \mu_{e}}\right)_{T}.
\end{align}
Since we have an analytical expression for the many-body T-matrix, we can also calculate the density analytically. By using similar techniques as shown in Ref.\,\cite{Henk}, we perform the summation over Matsubara frequencies by contour integration. Note that we have to be careful with the branch cut and the pole when choosing the contours. Ultimately, we find
\begin{align}\label{eq:chemint}
n = n_{\mathrm{id}} + n_{\mathrm{ex}} + n_{\mathrm{sc}},
\end{align}
where
\begin{align}\label{eq:chemintb}
n_{\mathrm{id}}&= \frac{1}{2 \pi^{2}} \left(\frac{2 m_{i}}{\hbar^{2}} \right)^{3/2} \int_{0}^{\infty} d\epsilon \, \sqrt{\epsilon}\,N_{\mathrm{FD}}(\epsilon - \mu_{i}) , \\ \nonumber
n_{\mathrm{ex}} &=\frac{\Theta(a)}{\pi^{2}} \left(\frac{2 (m_{e} + m_{h})}{\beta \hbar^{2}} \right)^{3/2} \int_{0}^{\infty} d\epsilon \, \sqrt{\epsilon} \, N_{\mathrm{BE}}(\epsilon_{\mathrm{ex}}), \\ \nonumber
n_{\mathrm{sc}} &=- \frac{1}{2 \pi^{3}} \frac{\Lambda_{\mathrm{th}}}{4 a \sqrt{\pi}} \left(\frac{2 (m_{e} + m_{h})}{\beta \hbar^{2}} \right)^{3/2} \\ \nonumber
&\times \int_{0}^{\infty} d\epsilon \int^{\infty}_{-\infty} dy \,  \sqrt{\epsilon}  \frac{N_{\mathrm{BE}}(\epsilon - \beta(\mu_{e} + \mu_{h}) + y^{2})}{y^{2} + \left(\Lambda_{\mathrm{th}}/4 a \sqrt{\pi} \right)^{2}},
\end{align}
with $\Theta$ the Heaviside step function, $\beta = 1 / k_{\mathrm{B}} T$, $\Lambda_{\mathrm{th}} = \left(2 \pi \hbar^{2} \beta / m_{r}\right)^{1/2}$ the thermal de Broglie wavelength, $N_{\mathrm{BE}}(x) = 1 / (e^{x} + 1 )$ the Bose-Einstein distribution function and we defined the dimensionless exciton energy $\epsilon_{\mathrm{ex}}$ as
\begin{align}\label{eq:exen}
\epsilon_{\mathrm{ex}} = \epsilon - \beta(\mu_{e} + \mu_{h}) - \frac{1}{16 \pi} \left(\frac{\Lambda_{\mathrm{th}}}{a} \right)^{2}.
\end{align} 
We see that there are three contributions to the carrier density. The first contribution is simply the ideal gas result. The second part originates from the exciton bound state and the third part describes the scattering continuum of electron-hole states. We can calculate the separate contributions to the density as a function of the value of the scattering length. In order to obtain these values, we first take a certain carrier density $n$. Then we can find the chemical potentials as a function of the scattering length, by using Eqs.\,\eqref{eq:chemint} and \,\eqref{eq:chemintb} for both $i = e$ and $i = h$. By resubstituting the chemical potentials, we obtain that every contribution to the density only depends on the scattering length and the carrier density. Therefore, we investigate the effect of changing the value of these physical parameters. Of course, because of screening the effective scattering length will also depend on the carrier density, but we ignore this in first instance and come back to this problem in a moment.
\newline
\indent In Fig.\,\ref{fig:BECBCS} we show the exciton contribution $n_{{\mathrm{ex}}}$ and the sum with the scattering contribution $n_{{\mathrm{sc}}}$ as a function of the scattering length. As an example, we take ZnO at room temperature and use $m_{e} = 0.28 m_{0}$, $m_{h} = 0.59 m_{0}$ with $m_{0}$ the bare electron mass, see Ref.\,\cite{Marijn}. First, we note from Eqs.\,\eqref{eq:chemintb} that there are only excitons for positive values of the scattering length and therefore we only show the number of excitons in this regime. For smaller values of the scattering length the contribution of the excitons becomes larger. Furthermore, by looking at the sum of the exciton and scattering contribution as a function of the interactions, we note that there is a smooth crossover from $\Lambda_{\mathrm{th}} / a \ll 0$ when the electrons behave as an ideal gas, to $\Lambda_{\mathrm{th}}/ a \gg 0$ where almost all electrons form excitons. Therefore, we note that this model for the semiconductor describes both the exciton regime and the regime where there is a electron-hole plasma. Moreover, we note that the sum of the exciton and the scattering contribution is an analog for the behavior of the Cooper-pair density in the BEC-BCS crossover \cite{BECBCS}. 

\subsection{Scattering length}
Up to now we have considered the scattering length as a free parameter. However, in a semiconductor the interaction depends on the value of the carrier density due to screening effets. Here we only consider the effects of static screening that can be described by using a Yukawa potential. Therefore, we can make our model selfconsistent by calculating the scattering length for the Yukawa potential.
\newline
\indent The standard procedure for calculating the scattering length is via the so-called Born series. However, for a Yukawa potential this will lead to divergences and taking into account only the first term of this expansion is not sufficient. Therefore, we here use a different approach. First, we note that the scattering length is associated with a low-energy two-body scattering wavefunction that is given by
\begin{align}\label{eq:wvsc}
\psi(r) \buildrel r \rightarrow \infty\over= A \left(1 - \frac{a}{r} \right),
\end{align}
where $A$ is a normalization constant that is usually taken to be 1 for many-body applications. Moreover, if we define $u(r) = r \psi(r)$ than this function satisfies the following radial Schr\"{o}dinger equation
\begin{align}\label{eq:rdse}
\frac{\hbar^{2}}{2 m_{r}} \frac{d^{2} u(r)}{dr^{2}} = V(r) u(r),
\end{align}
where $V(r)$ is a potential and $r$ is the relative coordinate between the electron and hole. Note that this is an equation for the relative motion between the electron and hole and therefore we have to use the reduced mass $m_{r}$ in the kinetic energy part. For relatively large distances the potential vanishes and therefore the wavefunction given by Eq.\,\eqref{eq:wvsc} is a solution to this equation. 
\begin{figure}[t]
\centerline{\includegraphics[scale=1]{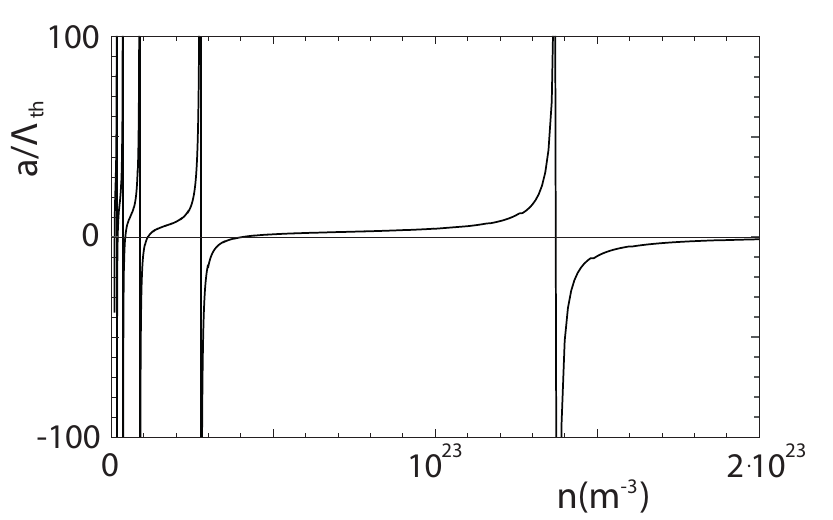}}
\caption{The scattering length for ZnO as a function of the carrier density by using the Yukawa potential with a screening length that is calculated for the ideal-gas parts of the electron and hole densities. We obtain that the last resonance occurs at the Mott-density $n_{\mathrm{M}} \simeq 2.3 \cdot 10^{24} \, \mathrm{m}^{-3}$.}
\label{fig:SLyuk}
\end{figure}
Hence, we have to solve this equation and look at the solution at large distances. To find the scattering length, we use that
\begin{align}
-a = \lim_{r \rightarrow \infty} \left(\frac{u(r)}{u'(r)} - r \right).
\end{align}
Thus by solving the radial Schr\"{o}dinger equation as written in Eq.\,\eqref{eq:rdse} we can find the scattering length.
\newline
\indent For the semiconductor we assume that there is only static screening and therefore we are allowed to take a Yukawa potential that is defined as
\begin{align}
V(r) = -\frac{e^{2}}{4 \pi \epsilon_{0} \epsilon_{r} r} e^{-r / \lambda_{s}},
\end{align}
with $e$ the electron charge, $\epsilon_{0}$ the vacuum permittivity, $\epsilon_{r}$ the appropriate relative dieletric constant and $\lambda_{s}$ the screening length. We calculate the screening length by assuming that only the unbound carriers screen the potential. Although this approximation leads to an overestimation of the screening length, we expect that this is a small effect since screening by bound carriers is weaker than screening by unbound carriers. 
\newline
\indent In this approximation the screening length $\lambda_{s}$ satisfies $\lambda^{-2}_{s} = \lambda^{-2}_{e} + \lambda^{-2}_{h}$ with
\begin{align}\label{eq:scideal}
\lambda_{i} = \sqrt{\frac{\epsilon_{0} \epsilon_{r}}{e^{2}}\frac{\partial \mu^{0}_{i}}{\partial n}},
\end{align}
the screening lengths of the electron and hole plasmas respectively. Note that in this formula the superscript $0$ of the chemical potential indicates that we calculate from the unbound carrier density by treating the electron and hole as ideal Fermi gases. To find the scattering length for the Yukawa potential we still need to specify the boundary conditions in Eq.\,\eqref{eq:rdse}. Since the potential is infinite at $r = 0$, we must take $u(0) = 0$. Moreover, we set $u^{\prime}(0) = 1$. Note that this latter boundary condition is simply a matter of normalization and does not affect the value of the scattering length.
\newline
\indent In Fig.\,\ref{fig:SLyuk} we display the scattering length as a function of the carrier density by determining the screening length with the ideal-gas parts of the electron and hole densities. For convenience we again consider ZnO with the values as stated in Ref.\,\cite{Marijn}. We observe that there are multiple resonances or equivalently values of the carrier density where the scattering length diverges. This is a consequence of the changing value of the screening length for different carrier densities. For larger carrier densities, the screening length becomes smaller and therefore roughly speaking the potential is less deep. Hence, for small carrier densities the Yukawa potential supports more bound states. Furthermore, every resonance corresponds to a bound state and we find that the last resonance occurs at the Mott-density $n_{\mathrm{M}} \simeq 2.3 \cdot 10^{24} \, \mathrm{m}^{-3}$. After that value of the carrier density, the scattering length always remains negative. Therefore, this calculation shows that there are only excitons for carrier densities that are smaller than the Mott-density. Note that this is in agreement with previously obtained results and the value for the Mott-density is within the range of published data \cite{Marijn}. 
\newline
\indent For large carrier densities we only have to take into account the last resonance. Moreover, for carrier densities that are close to $n_{M}$ we can approximate
\begin{align}\label{eq:approxa}
\frac{a(n)}{\Lambda_{\mathrm{th}}} \simeq 2.1 \frac{n_{\mathrm{M}}}{n_{\mathrm{M}} - n}.
\end{align} 

\subsection{Many-body effects}
We are primarly interested in the regime where the amount of excitons is negligible, because this is the regime where Bose-Einstein condensation of light is possible. From the results in the last section, we know that we therefore have to consider large carrier densities. At these high densities the many-body effects, i.e., the Fermi-Dirac distribution functions in Eq.\,\eqref{eq:TM}, become important. Therefore, we now focus on these contributions and we consider the part of the many-body T-matrix that has been neglected before, namely
\begin{align}
\frac{1}{(2 \pi)^{3}} \int\, d{\bf p}^{\prime} \frac{ N_{\mathrm{FD}}(\epsilon_{{\bf P} - {\bf p}^{\prime},e}) + N_{\mathrm{FD}}(\epsilon_{{\bf p}^{\prime},h})}{i \hbar \Omega_{n} - \epsilon_{{\bf P} - {\bf p}^{\prime},e} - \epsilon_{{\bf p}^{\prime},h} + \mu_{e} + \mu_{h}}.
\end{align}
In first instance we only consider the imaginary part and we neglect the real part. For further purposes we need to calculate this many-body T-matrix for $i \Omega_{n} \rightarrow \omega^{+} = \omega + i \epsilon$, where $\epsilon > 0$ is infinitesimally small. By switching to spherical coordinates and performing the angular integration, we find that there is only a non-zero imaginary part if
\begin{align}
y := \hbar \omega + \mu_{e} + \mu_{h} - \epsilon_{{\bf P}, e+h} > 0.
\end{align}
The imaginary part is given by
\begin{align}
\frac{\Theta  \left( y \right) }{4 \pi \hbar^{2} P} \left[ \int_{p^{h}_{\mathrm{min}}}^{p^{h}_{\mathrm{max}}} \, dp \, k \, m_{e} \, N_{\mathrm{FD}}(\epsilon_{{\bf p},h}) +   \langle e \leftrightarrow h\rangle \right],
\end{align}
where $P = |{\bf P}|$ and $\langle e \leftrightarrow h\rangle$ denotes that there is a similar contribution as the first integral where the indices $e$ and $h$ are interchanged. Furthermore,
\begin{align}
p^{h}_{\mathrm{min}} &= \frac{1}{\hbar} \sqrt{2 y \frac{m_{e} m_{h}}{m_{e} + m_{h}}} - \frac{m_{h}}{m_{e} + m_{h}} P, \\ \nonumber
p^{h}_{\mathrm{max}} &=  \frac{1}{\hbar} \sqrt{2 y \frac{m_{e} m_{h}}{m_{e} + m_{h}}} + \frac{m_{h}}{m_{e} + m_{h}} P,
\end{align}
where $p^{e}_{\mathrm{min}}$ and $p^{e}_{\mathrm{max}}$ are given by similar expression with again the indices $e$ and $h$ interchanged. 
\newline
\indent By using
\begin{align}
&\int_{p^{h}_{\mathrm{min}}}^{p^{h}_{\mathrm{max}}} \, dp \, p \, \, N_{\mathrm{FD}}(\epsilon_{{\bf p},h}) \\ \nonumber
&=-\frac{m_{h}}{\beta \hbar^{2}} \mathrm{Log}\left[ \frac{e^{-\beta \mu_{h}} + e^{-\beta \hbar^{2} (p^{h}_{\mathrm{max}})^{2}/ 2 m_{h}}}{e^{-\beta \mu_{h}} + e^{-\beta \hbar^{2} (p^{h}_{\mathrm{min}})^{2}/ 2 m_{h}}} \right],
\end{align}
we obtain for the full many-body T-matrix,
\begin{align}\label{eq:MBt}
&\frac{1}{T^{\mathrm{MB}}} = \frac{1}{V_{0}} - \Theta(-y) \frac{m_{r}^{3/2}}{\pi \hbar^{3}}  \sqrt{-\frac{y + i \epsilon}{2}} - \frac{i m_{e} m_{h}}{4 \pi \beta \hbar^{4} P} \\ \nonumber
&\times \Theta(y) \mathrm{Log}\left[\frac{\mathrm{cosh} \left(\frac{\beta}{2} \left\{\hbar \omega + C_{e + h} \right\}\right) + \mathrm{cosh} \left(\frac{\beta}{2} C_{e - h} \right)}{\mathrm{cosh} \left(\frac{\beta}{2} \left\{\hbar \omega - C_{e+h} \right\}\right) + \mathrm{cosh}\left(\frac{\beta}{2} C_{e-h} \right) } \right],
\end{align}
where
\begin{align}
C_{e + h} &= 4 \frac{\sqrt{m_{e} m_{h}}}{m_{e} + m_{h}}  \sqrt{y \, \epsilon_{{\bf P}, e+h}},  \\ \nonumber
C_{e - h} &= \mu_{e} - \mu_{h} + \frac{m_{e} - m_{h}}{m_{e} + m_{h}} \left(y - \epsilon_{{\bf P}, e+h} \right).
\end{align}
Note that this results includes the two-body result as given by Eq.\,\eqref{eq:2tm}. Moreover, in the context of ultracold Fermi gases a similar calculation has already been performed in Ref.\,\cite{Falco}. 
\begin{figure}[t]
\centerline{\includegraphics[scale=1]{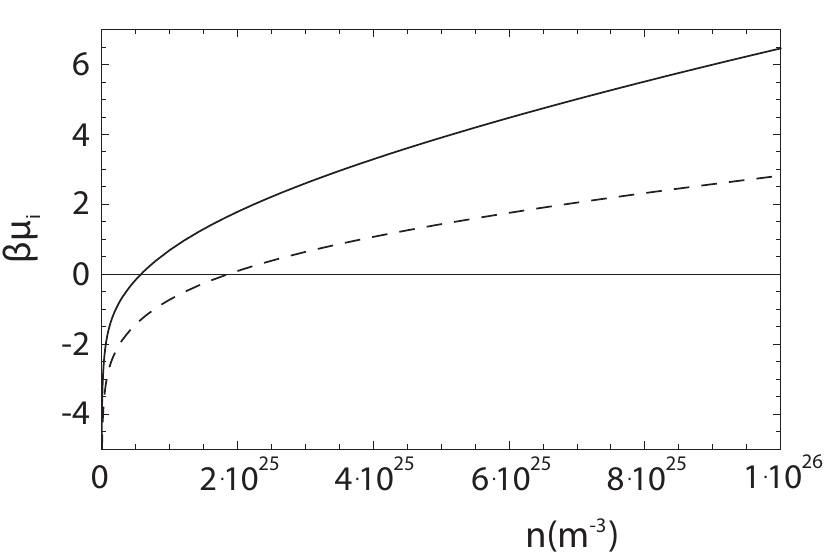}}
\caption{The chemical potential of the electrons $\mu_{e}$ (solid) and the chemical potential of the holes $\mu_{h}$ (dashed) for ZnO as a function of the carrier density by using the self-consistent scattering length for the Yukawa potential.}
\label{fig:ChemPot}
\end{figure}
As expected, we agree with that result if we put $m_{e} = m_{h} = m$ and $\mu_{e} = \mu_{h} = \mu$.
\newline
\indent Now we use this result to find a self-consistent solution for the chemical potential of the electrons and holes. Recall that
\begin{align}
n_{i} &=  -\frac{1}{V} \left(\frac{\partial \Omega}{\partial \mu_{i}}\right)_{T} = \frac{2}{V} \sum_{{\bf p}} N_{\mathrm{FD}}(\epsilon_{{\bf p},i}) \\ \nonumber
& - \frac{4 \hbar}{V} \sum_{{\bf P}} \int \, d({\hbar \omega}) \, \rho_{i}(\omega,{\bf P}) N_{\mathrm{BE}}(\omega),
\end{align}
where
\begin{align}
\rho_{i}(\omega,{\bf P}) = -\frac{1}{\pi \hbar} \mathrm{Im} \left[T^{\mathrm{MB}}(\omega, {\bf P}) \frac{\partial}{\partial \mu_{i}} \frac{1}{T^{\mathrm{MB}}(\omega, {\bf P})} \right].
\end{align}
As in the previous sections, we only want to consider optical excitation. Therefore, we for simplicity from now on neglect the dependence of the many-body T-matrix on $\mu_{e} - \mu_{h}$ and set
\begin{align}
C_{e - h} = \frac{m_{e} - m_{h}}{m_{e} + m_{h}} \left(y - \epsilon_{{\bf P}, e+h} \right).
\end{align}
In this case the spectral function is the same for $i = e$ and $i = h$ and we again define the carrier density as $n = n_{e} = n_{h}$. 
\newline
\indent The spectral function $\rho(\omega,{\bf P}) = \rho_{e}(\omega,{\bf P}) = \rho_{h}(\omega,{\bf P})$ of the pairs contains two different contributions that we can separate by considering either $y < 0$ or $y> 0$. For $y < 0$, we obtain that the Logarithm does not contribute in Eq.\,\eqref{eq:MBt}. In this case we can analytically determine the spectral function and we find
\begin{align}
\rho&(\omega,{\bf P}) \\ \nonumber
&= \frac{\Theta\left( a \right)}{\hbar} \delta \left(\hbar \omega -  \epsilon_{{\bf P},e + h} + \mu_{e} + \mu_{h} + \frac{\Lambda_{\mathrm{th}}^{2}}{16 \pi \beta a^{2}} \right).
\end{align}
By using the result of Eq.\,\eqref{eq:exen} we recognize that this contribution to the spectral function is simply a delta-function at the exciton energy. Again recall that this contribution is only present when the scattering length is positive and from the results in the previous section this implies that for carrier densities larger than the Mott density this contribution is not present. The other contribution to the spectral function is obtained by taking $y > 0$. In this case the square root in the expression for the many-body T-matrix vanishes and we are left with the contribution from the logarithm. This represents the scattering continuum of electrons and holes.
\newline
\indent In Fig.\,\ref{fig:ChemPot} we show the result for the chemical potentials of electrons and holes as a function of the carrier density $n$ for ZnO as discussed in Ref.\,\cite{Marijn}. In comparison with the calculation in that article, where the chemical potentials are determined by treating the electron and holes as two ideal Fermi gases, we find slightly smaller values for both chemical potentials. For example, we find for the value of the carrier density that corresponds to population inversion, i.e., $\mu_{e} + \mu_{h} > 0$, a value that is roughly two times larger than stated in Ref.\,\cite{Marijn}.

\section{Light in semiconductor microcavities}
\label{sec:Light}
Now that we worked out the model for the semiconductor, we consider the full situation and include the coupling to an external light field. Similar to Bose-Einstein condensation of light in a dye-filled cavity, we can describe the complete dynamics using the Schwinger-Keldysh formalism as presented in Ref.\,\cite{AW}. As will follow from the results in this section, in this case the self-energy as given in the Langevin field equation in Ref.\,\cite{AW} is proportional to the susceptibility of the semiconductor. However, also for Bose-Einstein condensation of light in nano-fabricated semiconductor microcavities the system relaxes towards a steady state that can be described by standard equilibrium methods. It is important to realize here that throughout our paper we are always considering the thermalized case, and in particular are not discussing the fully non-equilibrium laser regime of this system. We start from the action
\begin{align}\label{eq:action}
S[&a_{{\bf k}},a^{*}_{{\bf k}},\phi_{{\bf k}},\phi^{*}_{{\bf k}}] = S_{\mathrm{sc}}[\phi_{{\bf k}},\phi^{*}_{{\bf k}}] \\ \nonumber
+&\, \sum_{{\bf k}} \int_{0}^{\hbar \beta} d\tau \, a_{{\bf k}}^{*}(\tau) \left\{\hbar \frac{\partial}{\partial \tau} + \epsilon_{\mathrm{ph}}({\bf k}) - \mu_{\mathrm{ph}} \right\} a_{{\bf k}}(\tau) \\ \nonumber
-&\, \frac{g_{eh}}{\sqrt{V}} \sum_{{\bf k},{\bf p},\alpha} \int_{0}^{\hbar \beta} d\tau \, a^{*}_{{\bf k}}(\tau) \phi_{h,{\bf p},-\alpha}(\tau) \phi_{e,{\bf k} - {\bf p},\alpha}(\tau) \\ \nonumber
-&\, \frac{g_{eh}}{\sqrt{V}} \sum_{{\bf k},{\bf p},\alpha} \int_{0}^{\hbar \beta} d\tau \, a_{{\bf k}}(\tau) \phi^{*}_{e,{\bf k} - {\bf p},\alpha}(\tau) \phi^{*}_{h,{\bf p},-\alpha}(\tau).
\end{align}
This action contains several parts. First, $S_{\mathrm{sc}}[\phi_{{\bf k}},\phi^{*}_{{\bf k}}]$ describes the semiconductor, which is the momentum-space representation of Eq.\,\eqref{eq:acsm}. The second part describes the light with $a_{{\bf k}}^{*}(\tau)$ and $a_{{\bf k}}(\tau)$ the photon fields. Here, $\epsilon_{\mathrm{ph}}({\bf k})$ is the kinetic energy of the photons. By performing a full band structure calculation, e.g. see Ref.\,\cite{Sebas}, we know that
\begin{align}
\epsilon_{\mathrm{ph}}({\bf k}) \cong E_{0} + \frac{\hbar^{2} ({\bf k} - {\bf k}_{0})^{2}}{2 m},
\end{align}
with $E_{0}$ the energy of the photon at the minimum of the band with respect to the energy of the band gap of the semiconductor, ${\bf k}_{0}$ the wavenumber at the minimum of the band and $m$ the effective mass that arises from the local curvature of the band. In equilibrium, the sum of the number of electron and photons and the sum of holes and photons is conserved. Therefore, in equilibrium we have
\begin{align}
\mu_{\mathrm{ph}} = \mu_{e} + \mu_{h}.
\end{align} 
The last two terms of Eq.\,\eqref{eq:action} describe the interaction of the light with the semiconductor. The first of these terms corresponds to the annihilation of electron-hole pairs by the electric field and the second term is the creation of electron-hole pairs by the electric field. Recall that $\alpha$ represents the spin degeneracy of the electrons and holes and can be either $\uparrow$ or $\downarrow$. 
\begin{figure}[t]
\centerline{\includegraphics[scale=1]{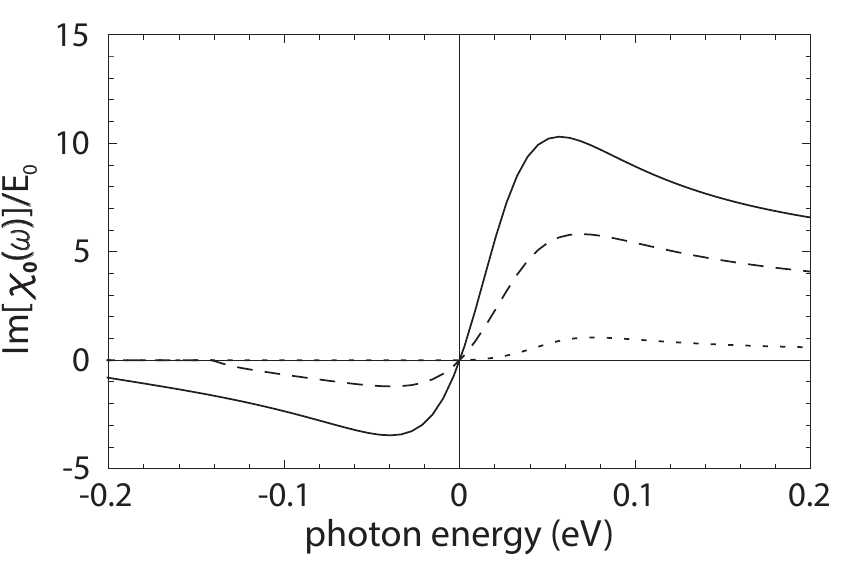}}
\caption{The imaginary part of the susceptibility $\chi_{{\bf 0}}(\omega)$ divided by the energy of the photon at the minimum of the band $E_{0}$ as a function of the photon energy for $g_{eh} \simeq 1.7 \cdot 10^{-32}\, \mathrm{J} \cdot \mathrm{m}^{3/2}$ and $E_{0} = 0.72\,\mathrm{eV}$. The dotted, dashed and solid curves corresponds to a carrier density of $n = 10^{25}\, \mathrm{m}^{-3}$, $n = 5 \cdot 10^{25}\, \mathrm{m}^{-3}$ and $n = 10^{26}\, \mathrm{m}^{-3}$ respectively.}
\label{fig:Suscep}
\end{figure}
In the interactions terms we only consider electrons and holes with opposite spin and therefore we only take into account transitions without spin-flip. 
\begin{figure*}[t!]
\centerline{\includegraphics[scale=1]{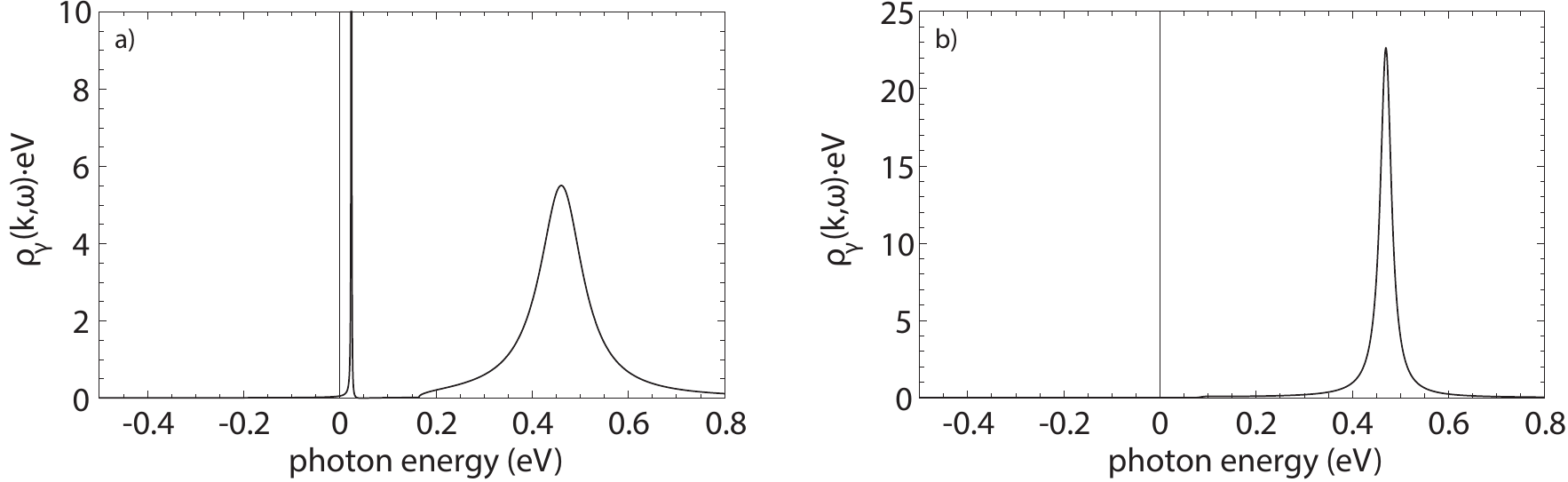}}
\caption{The spectral function of the photons above and below the Mott density of $n \simeq 2.3 \cdot 10^{24} \, \mathrm{m}^{-3}$ for $g_{eh} \simeq 9 \cdot 10^{-34}\, \mathrm{J} \cdot \mathrm{m}^{3/2}$ and $E_{0} = 0.4\,\mathrm{eV}$. In Figure a) the density is $n = 3.1 \cdot 10^{23} \, \mathrm{m}^{-3}$ and Figure b) is for a carrier density of $n = 1 \cdot 10^{25} \, \mathrm{m}^{-3}$. Below the Mott density there are two different contributions corresponding to the upper and lower exciton-polaritons. However, above the Mott density excitons do not exist and therefore there are only pure photons that acquire a finite lifetime through the interaction with the electrons and holes. This is the regime in which Bose-Einstein condensation of light is possible.}
\label{fig:SpecMott}
\end{figure*}
Moreover, we introduced the coupling constant $g_{eh}$ and we assumed that the coupling is independent of energy. Finally, note that the photons and the corresponding momentum ${\bf k}$ are two dimensional. However, the electron and holes are three dimensional and therefore ${\bf p}$ is also a three-dimensional vector. This convention will be used throughout the remainder of this paper.
\newline
\indent To obtain the effect of the coupling to the semiconductor on the behaviour of the photon gas, we use second order perturbation theory in the coupling constant $g_{eh}$. By following the same steps as in the appendix of Ref.\,\cite{Marijn}, we find the effective action for the photons to be
\begin{align}\label{eq:actionsuscep}
&S^{\mathrm{eff}}[a_{{\bf k}},a^{*}_{{\bf k}}] := - \sum_{{\bf k},n} \, a_{{\bf k},n}^{*} \hbar G_{\gamma}^{-1}({\bf k}, i \omega_{n}) a_{{\bf k},n} \\ \nonumber
&= \sum_{{\bf k},n} \, a_{{\bf k},n}^{*} \Big{\{}-i\hbar\omega_{n} + \epsilon_{\mathrm{ph}}({\bf k})  - \mu_{\mathrm{ph}} - \chi_{{\bf k}}(\omega_{n}) \Big{\}} a_{{\bf k},n}.
\end{align}
Here the susceptibility $\chi_{{\bf k}}(\omega_{n})$ acts as a selfenergy for the photons. In the Nozi\`{e}res-Schmitt-Rink approximation it is given by
\begin{align}
\chi_{{\bf k}}(\omega_{n}) = \frac{g^{2}_{eh}}{V} \sum_{{\bf p}} \chi_{{\bf k},{\bf p}}(\omega_{n}),
\end{align}
with
\begin{align}
\chi_{{\bf k},{\bf p}}(\omega_{n}) = \chi^{0}_{{\bf k},{\bf p}}(\omega_{n}) \left(1 - \frac{V_{0}}{V} \sum_{{\bf p}^{\prime}} \chi_{{\bf k},{\bf p}^{\prime}}(\omega_{n}) \right).
\end{align}
Hence,
\begin{align}
\chi_{{\bf k},{\bf p}}(\omega_{n}) = \frac{\chi^{0}_{{\bf k},{\bf p}}(\omega_{n})}{1 + \frac{V_{0}}{V} \sum_{{\bf p}^{\prime}} \chi^{0}_{{\bf k},{\bf p}^{\prime}}(\omega_{n})},
\end{align}
where
\begin{align}
\chi^{0}_{{\bf k},{\bf p}}(\omega_{n}) = \frac{1 - N_{\mathrm{FD}}(\epsilon_{{\bf k} - {\bf p},e}) - N_{\mathrm{FD}}(\epsilon_{{\bf p},h})}{i \hbar \omega_{n} - \epsilon_{{\bf k} - {\bf p},e} - \epsilon_{{\bf p},h} + \mu_{e} + \mu_{h}}.
\end{align}
\indent From the previous section we know the effective interaction strength $V_{0}$ and the chemical potentials as a function of the carrier density. Therefore, we can determine the finite lifetime effects of the photons due to the interaction with the electron-hole plasma by calculating the imaginary part of the susceptibility $\chi_{{\bf k}}(\omega)$ for every carrier density. In principle we can determine the susceptibility for arbitrary photon momentum ${\bf k}$. However, since we are primarily interested in Bose-Einstein condensation of the photons, we from now onwards only consider ${\bf k} = {\bf 0}$.
\newline
\indent In Fig.\,\ref{fig:Suscep} we display the imaginary part of the susceptibility for $g_{eh} \simeq 1.7 \cdot 10^{-32}\, \mathrm{J} \cdot \mathrm{m}^{3/2}$. Note that this values only changes the absolute value of the imaginary part and does not affect the qualitative behaviour. However, the actual value for these physical parameter should be obtained from experiments on optical spectra of the semiconductor. We find that for small carrier densities the imaginary part is only nonzero for positive photon energies. However, for carrier densities that are large enough, we also obtain negative contributions for negative energies. This is a consequence of the fact that only for large enough carrier densities there is gain of photons by emission processes. Moreover, we find that for large carrier densities, where there is substantial gain of photons, the imaginairy part of the susceptibility is linear for small values of the photon energy. Note that this is similar to the result for Bose-Einstein condensation of light in a dye-filled optical microcavity as presented in Ref.\,\cite{AW}. In this regime of large carrier densities we can combine the finite lifetime effects of the photons in a single dimensionless parameter $\alpha$, which is simply the slope of the imaginairy part of the susceptibility at the origin. Note that we considered a homogeneous semiconductor coupled to an external light field. However, in the proposed experiment as envisaged here, the semiconductor contains airholes, in which there are of course no electron or holes. In this case we need to multiply $\alpha$ by $1 - \eta_{\mathrm{air}}$, where $\eta_{\mathrm{air}}$ corresponds to the fraction of the volume of the holes of the semiconductor. 
\newline
\indent Up to now we only considered carrier densities that are larger than the Mott density. In this case we are in the pure photon regime, whereas for carrier densities that are smaller than the Mott density the photons are coupled to the exciton and therefore exciton-polaritons excitations exist. To emphasize the difference between the two density regimes, we now investigate the spectral function of the photons in the two different regimes. We define the spectral function of the photons as
\begin{align}
\rho_{\gamma}({\bf k},\omega) = \frac{-1}{\pi \hbar} \mathrm{Im} \left[G_{\gamma}({\bf k}, \omega^{+}) \right],
\end{align}
where $G_{\gamma}({\bf k}, i \omega_{n})$ is defined in Eq.\,\eqref{eq:actionsuscep} and $\omega^{+} = \omega + i \epsilon$ with $\epsilon$ infinitesimally small.
\newline
\indent In Fig.\,\ref{fig:SpecMott} we show the dimensionless spectral function $\rho_{\gamma}({\bf k},\omega) \cdot \mathrm{eV}$ for $g_{eh} \simeq 9 \cdot 10^{-34}\, \mathrm{J} \cdot \mathrm{m}^{3/2}$ and $E_{0} = 0.4\,\mathrm{eV}$ as a function of the photon energy. Below the Mott density we have two distinct contributions. Due to the coupling of the excitons to the photons, we have two peaks corresponding to the lower and upper exciton-polariton branches. Note that the photons that are part of the upper exciton-polariton acquire a finite lifetime, since their energy is larger than the energy threshold for the scattering continuum of electrons and holes. However, for carrier densities that are larger than the Mott density, the excitons are no longer present. In this case the spectral function only has a pure photon contribution, where the interaction with the semiconductor results into a finite lifetime of the photon. Therefore, in this spectral function we clearly find the physical differences between the exciton-polariton and the photon limits of the crossover. It is important to note that Bose-Einstein condensation of photons appears only in the regime described in Fig.\,\ref{fig:SpecMott}\,b.

\section{Scissors modes} 
\label{sec:Scissors}
In the previous section we constructed a model for Bose-Einstein condensation of light in nano-fabricated semiconductor microcavities. From now onwards we focus on an example of a many-body phenomenon of light that cannot be studied the currents experiments on Bose-Einstein condensation of photons, but can be investigated in the proposed experimental set-up. Namely, we consider probing the superfluidity of light via the excitation of scissor modes.
\newline
\indent We consider a two-dimensional photon gas with effective mass $m$ and an effective point-like interaction $g$ in an external harmonic trapping potential $V^{\mathrm{ext}}({\bf x})$ with trapping frequencies $\omega_{x}$ and $\omega_{y}$. Note that in the nano-fabricated semiconductor microcavities the harmonic potential arises by systematically increasing the size of the holes from the center to the edge of the semiconductor. Furthermore, in the previous section we showed that there is an excitation exchange between the photon gas and the electron-hole plasma, but, apart from such processes, photons are assumed to be conserved. Therefore, we are allowed to consider this quasi-equilibrium photon gas in the grand-canonical ensemble and to introduce the chemical potential $\mu$ of the photons. Moreover, we introduce a dimensionless interaction parameter $g$ that describes the interactions between the photons. Note that an explicit expression for the parameter can be obtained from Eq.\,\eqref{eq:action} by applying fourth order perturbation theory in the coupling constant $g_{eh}$. However, here we simply use a phenomenological approach to incorporate the interactions between the photons and leave the calculation of the photon-photon interaction for future work.
\newline
\indent We are interested in the dynamics of a condensate of light after a sudden rotation of the trap. In the following, we treat the local density and local phase of the condensate separately and we consider a condensate with a large number of photons such that we can use the Thomas-Fermi approximation. 
\begin{figure}[t]
\centerline{\includegraphics[scale=2.25]{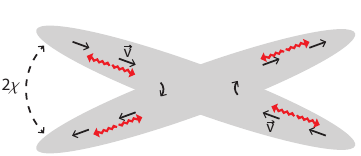}}
\caption{The cigar-shaped superfluid photon gas after a sudden rotation of the trapping potential by a small angle $\chi$. The dashed lines indicates the oscillations of the condensate and the solid arrows denote the irrotational velocity $\vec{v}$ of the condensed photons. The red wiggly arrows represent the decay products of the scissors mode quanta that can be used to demonstrate the dynamical Casimir effect.}
\label{fig:ScissorPRL}
\end{figure}
In this approximation the equilibrium condensate density $n_{0}({\bf x})$ is equal to
\begin{align}\label{eq:dens}
n_{0}({\bf x}) = \frac{\mu}{g} \left[1 - \left(x/R_{\mathrm{TF},x}\right)^{2} - \left(y/R_{\mathrm{TF},y}\right)^{2}  \right],
\end{align}
with $R^{2}_{\mathrm{TF},i} = 2 \mu / m \omega_{i}^{2}$ the Thomas-Fermi radius in the corresponding direction. Furthermore, the density is zero outside the ellipse that is spanned by these two radii. A rotation of the trap by a small angle $\chi$ results into a change in the condensate density $\delta n({\bf x})$ that is given by
\begin{align}\label{eq:scisdens}
\delta n({\bf x},t) := \left[n_{0}({\bf x}^{\prime}) -  n_{0}({\bf x})\right]e^{-i \omega t} = C x y e^{-i \omega t},
\end{align}
where ${\bf x}^{\prime}$ are the coordinates after the sudden rotation of the trap and $C := 2 \chi \mu (R_{\mathrm{TF},x}^{-2} - R_{\mathrm{TF},y}^{-2}) / g$. Here we used that for an eigenmode the time dependence is harmonic with angular frequency $\omega$. To obtain more information about the fluctuations of the local phase, we consider the hydrodynamic equations of the condensate as for example can be found in Ref.\,\cite{Smith}. By assuming that both the velocity of the condensate and the density fluctuations $\delta n ({\bf x})$ are small, we find for the phase of the condensate
\begin{align}\label{eq:scisphas}
\delta\phi({\bf x},t) = \frac{g}{i \hbar \omega} C x y e^{-i \omega t}.
\end{align}
Moreover, for a harmonic trapping potential the frequency of the scissors mode is given by $\omega = (\omega_{x}^{2} + \omega_{y}^{2})^{1/2}$.
\newline
\indent In the normal state the situation is different and we have to distinguish between the collisional and collisionless regime. In the collisional regime the frequency of the excitations is the same as in the superfluid phase \cite{SM1}. However, if the gas is dilute and the interactions are weak, the system is in the collisionless regime and the analog of the scissors-mode frequency is equal to $\omega_{x} + \omega_{y}$. By comparing the classical collision rate with the trap frequencies, we can make a distinction between both regimes. By using the expressions as specified in Ref.\,\cite{SM1}, we know how to construct the experiment conditions such that we are in the collisionless regime and we can distinguish whether the photons are superfluid or not by measuring the frequencies of the excitations of the photon gas after applying a rotation to the trap.

\subsection{Damping of scissors modes}
\label{sec:DampingScissors}
We have just seen that the superfluidity of the photons can be studied by looking at the frequencies of the excitations after applying a sudden rotation to the trap. These scissors modes result in oscillations that can directly be observed in experiments. However, due to the coupling of the photons with the electron-hole plasma it is worthwhile investigating how this coupling leads to the damping of these oscillations.
\newline
\indent We have demonstrated that for a Bose-Einstein condensate of light, at large carrier densities the effects of the electron-hole plasma can be characterized by a single dimensionless damping parameter $\alpha$ that depends on the external pumping. Moreover, the photon gas equilibrates to a steady state that is a dynamical balance between particle losses and external pumping. In the following we start from this steady state and we investigate the associated decay processes of the scissors modes. In particular, we show that the decay rate depend on the value of $\alpha$ and thereby on the external pumping.
\newline
\indent To observe properties of the damping, we are primarily interested in configurations that allow for many decay processes. For an elongated trap, the difference between the energies of two adjacent modes in the direction with the small trap frequency is small. Therefore, we expect that in that case the energy of the excitations in the long direction almost forms a continuum and the scissors mode quanta can decay into many other modes. Hence, this elongated trap is particularly interesting and we only consider this configuration throughout the remainder of this paper. A summary of the proposed structure is displayed in Fig.\,\ref{fig:ScissorPRL}.
\newline
\indent We are interested in damping processes where fluctuations of the condensate induce the creation of non-condensed excitations and therefore we only have to consider the interaction part of the hamiltonian. We substitute for the creation and annihilation operator of the photons $\hat{\psi}({\bf x},t) = \langle \hat{\psi}({\bf x},t) \rangle + \delta \hat{\psi}({\bf x},t)$ and we only consider terms up to quadratic order in the fluctuations $\delta \hat{\psi}({\bf x},t)$. In this Bogoliubov approximation, we therefore study
\begin{align}
\hat{H}_{\mathrm{int}}(t) &= \frac{g}{2} \int d{\bf x} \, \langle \hat{\psi}({\bf x},t) \rangle^{2} (\delta \hat{\psi}^{\dagger}({\bf x},t) )^{2}  \\ \nonumber
&+ \frac{g}{2} \int d{\bf x} \, \langle \hat{\psi}^{\dagger}({\bf x},t)  \rangle^{2} (\delta \hat{\psi}({\bf x},t) )^{2} \\ \nonumber
&+ g  \int d{\bf x} \, |\langle \hat{\psi}({\bf x},t) \rangle|^{2} \delta \hat{\psi}^{\dagger}({\bf x},t)  \delta \hat{\psi}({\bf x},t).
\end{align}
Now we explicitly separate the dynamics of the local phase and density of the condensate by writing $\langle \hat{\psi}({\bf x},t) \rangle = \sqrt{n_{0}({\bf x}) + \delta n({\bf x},t) }e^{i \delta\phi({\bf x},t)}$, with $n_{0}({\bf x})$ the equilibrium condensate density and both $\delta n({\bf x},t)$ and $\delta\phi({\bf x},t)$ fluctuations that are known from the calculations in the previous section. Since these fluctuations are small, we only consider the first non-vanishing term in the condensate fluctuations. For the harmonic potential that is considered here, the parts of the hamiltonian that are linear in the condensate fluctuations $\delta n({\bf x},t)$ and $\delta\phi({\bf x},t)$ vanish, since the fluctuations are odd under $y \rightarrow -y$. Hence, there is no decay of a single scissors mode quantum at this level of approximation and we have to consider the parts of the hamiltonian that are quadratic in the condensate fluctuations.
\newline
\indent We introduce the Bogoliubov amplitudes $u_{j}({\bf x})$ and $v_{j}({\bf x})$ according to
\begin{align}\label{eq:Bogamp}
\delta \hat{\psi}({\bf x},t) &= \sum_{j} \left[u_{j}({\bf x}) \hat{b}_{j}(t) - v^{*}_{j}({\bf x}) \hat{b}^{\dagger}_{j}(t) \right], \\ \nonumber
\delta \hat{\psi}^{\dagger}({\bf x},t) &= \sum_{j} \left[u_{j}^{*}({\bf x}) \hat{b}^{\dagger}_{j}(t)  - v_{j}({\bf x}) \hat{b}_{j}(t) \right],
\end{align}
where $j = 1,2,...$ indicates the mode number of the excitation. Note that we neglect the contributions coming from $j = 0$, since we are only interested in the decay into non-condensed excitations and the $j = 0$ term correspond to the dynamics of the global phase of the condensate. Therefore, we have
\begin{align}\label{eq:hamfer}
\hat{H}&_{I}(t) = e^{-2 i \omega t} \sum_{j,j^{\prime} \neq 0} H_{j,j^{\prime}} \hat{b}^{\dagger}_{j}(t) \hat{b}^{\dagger}_{j^{\prime}}(t) \\ \nonumber
&= \frac{(g C)^{2}}{\hbar \omega} e^{-2 i \omega t} \sum_{j,j^{\prime} \neq 0} \left(H^{1}_{j,j^{\prime}} + \frac{g H^{2}_{j,j^{\prime}}}{\hbar \omega}\right) \hat{b}^{\dagger}_{j}(t) \hat{b}^{\dagger}_{j^{\prime}}(t),
\end{align}
where 
\begin{align}\label{eq:hamfer2}
H^{1}_{j,j^{\prime}} &=  \int d{\bf x} \, x^{2}y^{2} \left[u_{j}^{*}({\bf x}) u_{j^{\prime}}^{*}({\bf x}) - v_{j}^{*}({\bf x}) v_{j^{\prime}}^{*}({\bf x}) \right], \\ \nonumber
H^{2}_{j,j^{\prime}} &=  \int d{\bf x} \, x^{2}y^{2} n_{0}({\bf x}) \left[u_{j}^{*}({\bf x}) u_{j^{\prime}}^{*}({\bf x}) + v_{j}^{*}({\bf x}) v_{j^{\prime}}^{*}({\bf x}) \right]. 
\end{align}
To make further progress, we need to determine the Bogoliubov amplitudes and determine the time dependence of the operators $\hat{b}_{j}(t)$ and $\hat{b}^{\dagger}_{j}(t)$. Hence, we need to solve the Bogoliubov-de Gennes equations.
\newline
\indent In two dimensions the general solution of the Bogoliubov-de Gennes equations is not known. However, in the elongated configuration we expect that the dynamics of the long-wavelength photons is approximately one dimensional. Therefore, we use the exact solution of the Bogoliubov-de Gennes equations in one dimension, see e.g. Ref.\,\cite{Wal}, in order to make a proper variational ansatz for the form of the non-condensed fluctuations in the elongated configuration. Thus, we consider the ansatz
\begin{align}\label{eq:ansatz}
u_{j}({\bf z}) &= \frac{1}{2 \sqrt{C_{j}}} \left[\alpha_{j} \sqrt{1 - |{\bf z}|^{2}} + \frac{\beta_{j}}{\sqrt{1-  |{\bf z}|^{2}}} \right] P_{j}(\tilde{x}), \\ \nonumber
v_{j}({\bf z}) &= \frac{1}{2 \sqrt{C_{j}}} \left[\alpha_{j} \sqrt{1 - |{\bf z}|^{2}} - \frac{\beta_{j}}{\sqrt{1 -  |{\bf z}|^{2}}} \right] P_{j}(\tilde{x}),
\end{align}
with ${\bf z} = (\tilde{x},\tilde{y}) = (x/R_{\mathrm{TF},x},y/R_{\mathrm{TF},y})$ and $ P_{j}(\tilde{x})$ the j-th Legendre polynomial. Furthermore, the constants $\alpha_{j}$ and $\beta_{j}$ are given by
\begin{align}\label{eq:const}
\alpha_{j} &= \frac{1}{\sqrt{R_{\mathrm{TF},x} R_{\mathrm{TF},y}}} \sqrt{\frac{\mu}{\hbar \Omega_{j}}}, \\ \nonumber
\beta_{j} &= \frac{1}{2 \sqrt{R_{\mathrm{TF},x} R_{\mathrm{TF},y}}} \sqrt{\frac{\hbar \Omega_{j}}{\mu}},
\end{align}
with $\Omega_{j} =  \omega_{x} \sqrt{j(j+1)/2}$ the energy eigenvalues of the one dimensional problem. Also $R_{\mathrm{TF},x}$ and $R_{\mathrm{TF},y}$ correspond to the Thomas-Fermi radius of the condensate in the specified direction. Moreover, we defined the constant $C_{j}$ which is given by
\begin{align}\label{eq:const2}
C_{j} = \int_{-1}^{1} d\tilde{x}\,(1 - \tilde{x}^{2})^{1/2} \left[P_{j}(\tilde{x})\right]^{2}.
\end{align}
This ansatz corresponds to a solution for an elongated trap where the excitation only propagates in the direction with the small trapping frequency. Note that if from the start we would have removed the $y$-dependence, our ansatz simplifies to the exact solution in one dimension as given in Ref.\,\cite{Wal}. 
\newline
\indent Since the photons are in a good first approximation equivalent to non-relativistic particles with an effective mass $m$ and point-like interaction with strength $g$, we consider the following action within the functional-integral formalism in the Bogoliubov approximation
\begin{align}
S&[\phi^{*},\phi] = \int d\tau \, \int d{\bf x}  \, \phi^{*}({\bf x},\tau) G^{-1}({\bf x},\tau) \phi({\bf x},\tau) \, \\ \nonumber
&+ \frac{g}{2} \int d\tau \,\int d{\bf x} \,  n_{0}({\bf x}) \left\{[\phi^{*}({\bf x},\tau)]^{2} + [\phi({\bf x},\tau)]^{2} \right\}.
\end{align}
Here we defined
\begin{equation}
G^{-1}({\bf x},\tau) = \hbar \frac{\partial}{\partial \tau} - \frac{\hbar^{2} \nabla^{2}}{2m} + V^{\mathrm{ext}}({\bf x}) - \mu + 2 g n({\bf x}).
\end{equation} 
Furthermore, $\phi({\bf x},\tau)$ and $\phi^{*}({\bf x},\tau)$ are the fields that describe the fluctuations that originate from the Bogliubov approximation $\psi({\bf x},\tau) = \langle \psi ({\bf x},\tau) \rangle + \phi({\bf x},\tau)$, $n_{0}({\bf x})$ is the condensate density, $n({\bf x})$ equals the total density and $V^{\mathrm{ext}}({\bf x})$ corresponds to the external trapping potential. We consider the usual Bogoliubov transformation with
\begin{align}
\phi({\bf x}, \tau) = \sum_{j} \left[u_{j}({\bf x}) b_{j}(\tau) - v^{*}_{j}({\bf x}) b^{*}_{j} (\tau) \right],
\end{align}
and the ansatz from Eq.\,\eqref{eq:ansatz} for the coherence factors. This allows us to rewrite
\begin{align}
S = \frac{1}{2} &\int d\tau \sum_{j,j^{\prime}} \left [ \begin{array} {c}
b_{j}(\tau) \\
b^{*}_{j}(\tau)
\end{array} \right ]^{\dagger} \cdot \Bigg{(}\left [ \begin{array} {cc}
G_{j,j^{\prime},11} & G_{j,j^{\prime},12} \\
G_{j,j^{\prime},12} & G_{j,j^{\prime},11}
\end{array} \right ]  \\ \nonumber
&+ \hbar \frac{\partial}{\partial \tau} \left [ \begin{array} {cc}
G_{j,j^{\prime}} & 0 \\
0 & - G_{j,j^{\prime}}
\end{array} \right ] \Bigg{)} \cdot
\left [ \begin{array} {c}
b_{j^{\prime}}(\tau) \\
b^{*}_{j^{\prime}}(\tau)
\end{array} \right],
\end{align}
with
\begin{align}
G_{j,j^{\prime}} &= \int d{\bf x} \, \left[u_{j}({\bf x}) u_{j^{\prime}}({\bf x}) - v_{j}({\bf x}) v_{j^{\prime}}({\bf x}) \right],\\ \nonumber
G_{j,j^{\prime},11} &= - g \int d{\bf x} \, n_{0}({\bf x}) \left[u_{j}({\bf x}) v_{j^{\prime}}({\bf x}) + u_{j^{\prime}}({\bf x}) v_{j}({\bf x}) \right]  \\ \nonumber
+ \int& d{\bf x} \, \left[u_{j}({\bf x}) G^{-1}({\bf x}) u_{j^{\prime}}({\bf x}) + v_{j}({\bf x}) G^{-1}({\bf x}) v_{j^{\prime}}({\bf x})\right],\\ \nonumber
G_{j,j^{\prime},12} &= g \int d{\bf x} \,  n_{0}({\bf x}) \left[u_{j}({\bf x}) u_{j^{\prime}}({\bf x}) + v_{j}({\bf x}) v_{j^{\prime}}({\bf x}) \right] \\ \nonumber
- \int& d{\bf x} \, \left[u_{j}({\bf x}) G^{-1}({\bf x}) v_{j^{\prime}}({\bf x}) + v_{j}({\bf x}) G^{-1}({\bf x}) u_{j^{\prime}}({\bf x}) \right].
\end{align}
Here we used that $u_{n}({\bf x})$ and $v_{n}({\bf x})$ are real. Furthermore,
\begin{equation}
G^{-1}({\bf x}) = G^{-1}({\bf x},\tau) - \hbar \frac{\partial}{\partial \tau} =  -\frac{\hbar^{2}}{2m} \frac{\partial^{2}}{\partial x^{2}} + g n_{0}({\bf x}),
\end{equation}
where we used the Gross-Pitaevski equation in the Thomas-Fermi limit. 
\begin{figure}[t]
\centerline{\includegraphics[scale=1]{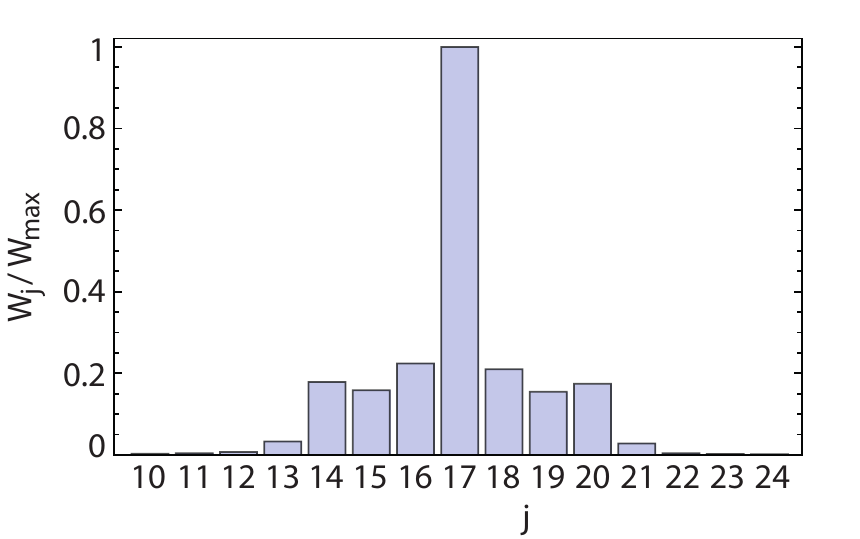}}
\caption{The decay rate of two scissors mode quanta into a non-condensed mode $W_{j} = \sum_{j^{\prime}} W_{j,j^{\prime}} (1 + \delta_{j,j^{\prime}}) $ with mode number $j$ for $\mu = 10\,\omega_{y}$, $\omega_{y} = 10\,\omega_{x}$ and $\alpha = 10^{-2}$. }
\label{fig:Goldenrule}
\end{figure}
Moreover, in this limit we can also neglect the second-order derivative with respect to $y$. This is valid as $G^{-1}$  acts on $u_{n}({\bf x})$ and $v_{n}({\bf x})$ that only have $y$-dependence in the density. 
\newline
\indent We consider $F_{j}({\bf x}) = u_{j}({\bf x}) + v_{j}({\bf x})$ and $G_{j}({\bf x}) = u_{j}({\bf x}) - v_{j}({\bf x})$ with $u_{j}({\bf x})$ and $v_{j}({\bf x})$ as given by Eq.\,\eqref{eq:ansatz}. Then,
\begin{align}
-\frac{\hbar^{2}}{2m} &\frac{\partial^{2}}{\partial x^{2}} u_{j}({\bf x}) =  -\frac{\hbar^{2}}{4m} \frac{\partial^{2}}{\partial x^{2}} (F_{j}({\bf x}) + G_{j}({\bf x})) \\ \nonumber
&=  -\frac{\hbar^{2}}{4m} \frac{\partial^{2}}{\partial x^{2}} F_{j}({\bf x}) = \frac{1-\tilde{x}^{2} - \tilde{y}^{2}}{1-\tilde{x}^{2}} \frac{\hbar \Omega_{j}}{2} G_{j}({\bf x}), 
\end{align}
where we used the results of Ref.\,\cite{Henk} and the fact that we are in the Thomas-Fermi limit and therefore neglected the derivatives of the densities. Namely, all these derivatives result into second-order derivatives of densities or terms that contain $n_{0}^{\prime}({\bf x}) / n_{0}({\bf x})$, which are all small in the Thomas-Fermi limit. Similarly, we find that
\begin{align}
-\frac{\hbar^{2}}{2m} &\frac{\partial^{2}}{\partial x^{2}} v_{j}({\bf x}) =  \frac{1-\tilde{x}^{2} - \tilde{y}^{2}}{1-\tilde{x}^{2}} \frac{\hbar \Omega_{j}}{2} G_{j}({\bf x}).
\end{align}
Now, we define
\begin{align}
I_{j,j^{\prime}} = \frac{1}{\sqrt{C_{j} C_{j^{\prime}}}}\int_{-1}^{1} d\tilde{x} \,  P_{j}(\tilde{x}) P_{j^{\prime}}(\tilde{x}) \sqrt{1 - \tilde{x}^{2}}.
\end{align}
Then,
\begin{align}
G_{j,j^{\prime}} &= \frac{1}{2} \left[\sqrt[4]{\frac{j^{\prime} (j^{\prime} + 1)}{j (j + 1)}}+  \sqrt[4]{\frac{j (j + 1)}{j^{\prime} (j^{\prime} + 1)}} \right] I_{j,j^{\prime}},\\ \nonumber 
G_{j,j^{\prime},11} &= \left(\frac{\hbar \Omega_{j^{\prime}}}{3} + \frac{\hbar \Omega_{j}}{2} \right) \sqrt[4]{\frac{j^{\prime} (j^{\prime} + 1)}{j (j + 1)}} I_{j,j^{\prime}},\\ \nonumber 
G_{j,j^{\prime},12} &= \left(\frac{\hbar \Omega_{j}}{2} - \frac{\hbar \Omega_{j^{\prime}}}{3} \right)  \sqrt[4]{\frac{j^{\prime} (j^{\prime} + 1)}{j (j + 1)}} I_{j,j^{\prime}}.
\end{align}
For now we ignore the coupling between modes where $j \neq j^{\prime}$. Note that this is in general a good approximation as the integral $I_{j,j^{\prime}}$ has a maximal value for $j = j^{\prime}$ and gradually decreases if $j$ is further and further away from $j^{\prime}$. This allows us to rewrite
\begin{align}
S &=  \frac{1}{2} \int d\tau \sum_{j} b_{j}^{*}(\tau) \left(\hbar \frac{\partial}{\partial \tau} + \frac{5 \hbar \Omega_{j}}{6} \right) b_{j}(\tau) \\ \nonumber
&-  \frac{1}{2} \int d\tau \sum_{j} b_{j}(\tau) \left(\hbar \frac{\partial}{\partial \tau} - \frac{5 \hbar \Omega_{j}}{6} \right) b^{*}_{j}(\tau) \\ \nonumber
&+\frac{\hbar \Omega_{j}}{12} \int d\tau \sum_{j} (b_{j}(\tau))^{2} +  (b^{*}_{j}(\tau))^{2}
\end{align}
\indent To complete our variational approach we determine the equations of motion, which allows us to determine the time dependence of the operators $b_{j}(\tau)$ and $b_{j}^{*}(\tau)$. We find for the equations of motion 
\begin{align}\label{eq:eomvar}
\left(\hbar \frac{\partial}{\partial \tau} + \frac{5 \hbar \Omega_{j}}{6} \right) b_{j}(\tau) + \frac{\hbar \Omega_{j}}{6} b^{*}_{j}(\tau) &= 0, \\ \nonumber
\left(\hbar \frac{\partial}{\partial \tau} - \frac{5 \hbar \Omega_{j}}{6} \right) b^{*}_{j}(\tau) - \frac{\hbar \Omega_{j}}{6} b_{j}(\tau) &= 0,
\end{align}
for every $j=0,1,...$. Hence,
\begin{align}\label{eq:bogvar}
b_{j}(\tau) &= c^{*}_{n} e^{\sqrt{\frac{2}{3}} \Omega_{j} \tau} + d_{j} e^{-\sqrt{\frac{2}{3}} \Omega_{j} \tau}, \\ \nonumber
b^{*}_{j}(\tau) &= c_{n} e^{-\sqrt{\frac{2}{3}} \Omega_{j} \tau} + d^{*}_{j} e^{\sqrt{\frac{2}{3}} \Omega_{j} \tau}.
\end{align}
Thus the energy of the excitations for $j=0,1,2,...$ is equal to
\begin{align}\label{eq:varen}
\hbar \omega_{j} = \hbar \Omega_{j}  \sqrt{2/3} = \hbar \omega_{x}\sqrt{j (j + 1)/3} .
\end{align}
\indent To quantify the effect of neglecting the non-diagonal terms, we numerically determine the eigenvalues of the full problem, where we consider terms from $j =1$ to $j  = 75$. The largest error of roughly $15$ percent occurs for $j=1$. For the other values of $j$ the error is even smaller and less than $10$ percent. Therefore, it is indeed a good first approximation to neglect the non-diagonal terms. Furthermore, by resubstituting the expressions in Eqs.\,\eqref{eq:bogvar} into Eqs.\,\eqref{eq:eomvar} we obtain
\begin{align}\label{eq:bogvar2}
d_{j}^{*} &= -\left(5 + 2 \sqrt{6} \right) c_{j}^{*}, \\ \nonumber
c_{j} &= \left(2 \sqrt{6} - 5 \right) d_{j}.
\end{align}
Hence, up to an overal normalization constant
\begin{align}
\phi({\bf x}, \tau) = \sum_{j} \left[\tilde{u}_{j}({\bf x}) d_{j} e^{-\omega_{j} \tau} - \tilde{v}_{j}^{*}({\bf x}) d^{*}_{j} e^{\omega_{j} \tau} \right],
\end{align}
where
\begin{align}
\tilde{u}_{j}({\bf x}) &= u_{j}({\bf x}) + \left(5 - 2 \sqrt{6} \right) v^{*}_{j}({\bf x}), \\ \nonumber
\tilde{v}^{*}_{j}({\bf x}) &= v^{*}_{j}({\bf x}) + \left(5 - 2 \sqrt{6} \right) u_{j}({\bf x}).
\end{align}
\indent By using these expressions for the Bogoliubov amplitudes, we find for $H_{j,j^{\prime}}$ as defined in Eqs.\,\eqref{eq:hamfer} and \eqref{eq:hamfer2},
\begin{align}
H_{j,j^{\prime}} &= \frac{2 \chi^{2} \mu^{2}}{3 \hbar \omega} \frac{\left(R^{2}_{\mathrm{TF},y} - R^{2}_{\mathrm{TF},x} \right)^{2}}{R^{2}_{\mathrm{TF},y} R^{2}_{\mathrm{TF},x}} \\ \nonumber
&\times \Bigg{[} \left(\sqrt{\frac{\omega_{j^{\prime}}}{\omega_{j}}} + \sqrt{\frac{\omega_{j}}{\omega_{j^{\prime}}}} +  \frac{\sqrt{\omega_{j} \omega_{j^{\prime}}}}{2 \omega} \right)  \frac{Z^{2,3}_{j,j^{\prime}}}{\sqrt{Z^{0,1}_{j,j} Z^{0,1}_{j^{\prime},j^{\prime}}}} \\ \nonumber
&+ \frac{16 \mu^{2}}{35 \hbar \omega \sqrt{\hbar \omega_{j} \hbar \omega_{j^{\prime}}}} \frac{Z^{2,7}_{j,j^{\prime}}}{\sqrt{Z^{0,1}_{j,j} Z^{0,1}_{j^{\prime},j^{\prime}}}}  \Bigg{]},
\end{align}
where we used the shorthand notation
\begin{align}
Z^{m,n}_{j,j^{\prime}} = \int_{-1}^{1} d\tilde{x} \, \tilde{x}^{m} \left(1 - \tilde{x}^{2} \right)^{n/2} P_{j}(\tilde{x}) P_{j^{\prime}}(\tilde{x}).
\end{align}
\newline
\indent To investigate the damping of the scissors modes, we are interested in the transition rate for creating two excitations with frequency $\omega_{j}$ and $\omega_{j^{\prime}}$ from the vacuum through the decay of two scissors mode quanta. By applying Fermi's Golden Rule, we obtain
\begin{eqnarray}\label{eq:probnoalpha}
W_{j,j^{\prime}} \simeq \frac{8 \pi |H_{j,j^{\prime}}|^{2}}{\hbar (1 + \delta_{j,j^{\prime}})} \rho(\omega_{j} + \omega_{j^{\prime}}),
\end{eqnarray}
where 
\begin{eqnarray}
\rho(\omega_{j} + \omega_{j^{\prime}}) = \frac{1}{\pi \hbar}\frac{\alpha (\omega_{j} + \omega_{j^{\prime}})}{(\omega_{j} + \omega_{j^{\prime}} - 2 \omega)^{2} + [\alpha (\omega_{j} + \omega_{j^{\prime}})]^{2}}.\,\,\,\,\,
\end{eqnarray}
Here we introduced a final density of states $\rho(E)$ to incorporate that, due to the interaction with the molecules, there is a probability that the photon is in a state with an energy that is within a small band around $\omega_{j} + \omega_{j^{\prime}}$. Note that for the current experiments we have that $\beta (\hbar \omega_{j} + \hbar \omega_{j^{\prime}}) \ll 1$ with $\beta = 1 / k_{\mathrm{B}} T$ the inverse of the thermal energy. Therefore, we used the low-energy approximation for the density of states \cite{AW}. As the dimensionless damping parameter $\alpha$ is small, we directly satisfy $W_{j,j^{\prime}} \simeq 0$ to a very good approximation if the photons scatter into a state with energy outside the small band around $\omega_{j} + \omega_{j^{\prime}}$. Note that the nonzero value of $\alpha$ makes our calculation specific to dissipative Bose-Einstein condensates and not immediately applicable to a cold atomic gas, where $\alpha = 0$ and Beliaev damping of the scissors modes is only possible in the presence of fine-tuned degeneracies \cite{Bel}. 
\newline
\indent In Fig.\,\ref{fig:Goldenrule} we show the decay rate $W_{j} = \sum_{j^{\prime}} W_{j,j^{\prime}} (1 + \delta_{j,j^{\prime}})$ for $\mu = 10\,\omega_{y}$, $\omega_{y} = 10\,\omega_{x}$ and $\alpha = 10^{-2}$. Because the frequency of the scissors mode $\omega$ is roughly equal to $\omega_{17}$, we obtain a peak for $j = 17$. Furthermore, we find that the decay of the scissors modes indeed leads to the population of several non-condensed modes.

\subsection{Density-density correlation function} 
\label{sec:DDcor}
\begin{figure}[t]
\centerline{\includegraphics[scale=1]{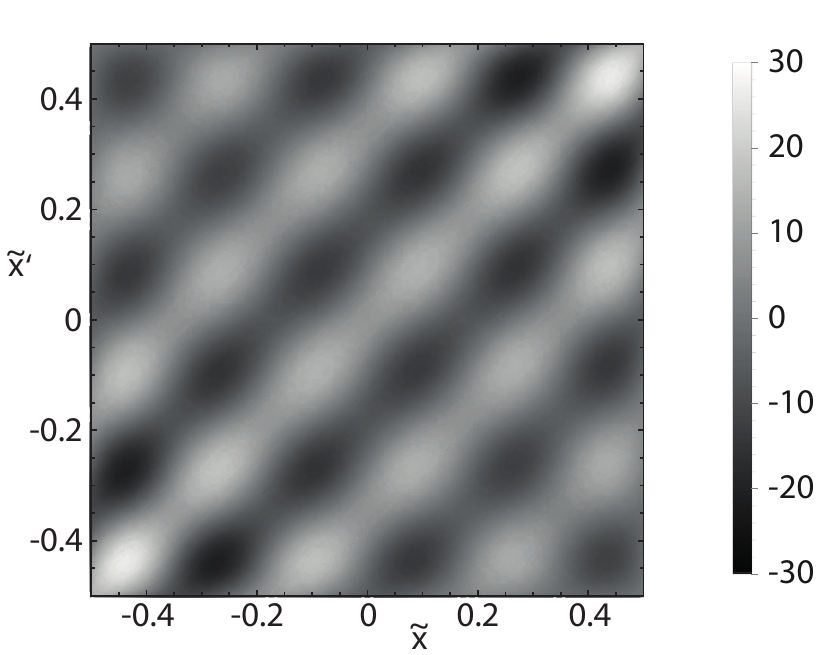}}
\caption{The density-density correlation function of the Bogoliubov excitations $8 n_{0} \xi^{2} g^{(2)}({\bf x},{\bf x}^{\prime},t)$, where $n_{0} \xi^{2} = \hbar^{2} / 2 m g$ with $\xi$ the so-called coherence length and $n_{0}$ the condensate density in the center of the trap, in terms of $(\hbar \omega_{x})^{2} \hbar \omega_{y} / \mu^{3}$ for $y = y^{\prime} = 0$.}
\label{fig:DDcor}
\end{figure}

We now consider the situation that the scissors modes are being excited and we consider the density-density correlation function in the operator formalism. Thus, we define
\begin{align}
g^{(2)}({\bf x},{\bf x}^{\prime},t) = \frac{\langle \hat{\rho}({\bf x},t) \hat{\rho}({\bf x}^{\prime},t) \rangle}{\langle \hat{\rho}({\bf x},t) \rangle \langle \hat{\rho}({\bf x}^{\prime},t) \rangle} - 1,
\end{align}
where $\hat{\rho}({\bf x},t) = \hat{\psi}^{\dagger}({\bf x},t)  \hat{\psi}({\bf x},t)$ is the density operator. We again take $\hat{\psi}({\bf x},t) = \langle \hat{\psi}({\bf x},t) \rangle + \delta \hat{\psi}({\bf x},t)$ and we explicitly separate the fluctuations that are described as scissors modes and Bogoliubov excitations by writing $\delta\hat{\psi}({\bf x},t) = \delta \hat{\psi}_{s}({\bf x},t) + \delta \hat{\psi}_{B}({\bf x},t)$. As a consequence, the density-density correlation function contains the density-density correlations from the scissors modes and also a term from the density-density correlations between the Bogoliubov modes. From now onwards we take $y = y^{\prime} = 0$ such that the contribution of the scissors modes vanishes, as can be seen explicitly in Eqs.\,\eqref{eq:scisdens} and \eqref{eq:scisphas}. 
\newline
\indent For the part with the Bogoliubov excitations we can use Eq.\,\eqref{eq:Bogamp} to rewrite the result in terms of the Bogoliubov amplitudes $u({\bf x})$ and $u({\bf x})$. We find
\begin{align}\label{eq:dendenback}
g^{(2)}({\bf x},&{\bf x}^{\prime},t) = \frac{1}{\sqrt{n_{0}({\bf x}) n_{0}({\bf x}^{\prime})}} \sum_{n} \left(u_{n}({\bf x}) - v_{n}({\bf x})\right)  \\ \nonumber
&\times \left(u_{n}({\bf x}^{\prime}) - v_{n}({\bf x}^{\prime})\right) \left\{1 + 2 \left \langle \hat{b}_{n}^{\dagger}(t) \hat{b}_{n}(t) \right\rangle \right\},
\end{align}
where $N_{n}(t) = \left \langle \hat{b}_{n}^{\dagger}(t) \hat{b}_{n}(t) \right\rangle$ is the number of excitations in a mode $n$ at time $t$. The contribution of $g^{(2)}({\bf x},{\bf x}^{\prime},t)$ that is independent of $N_{n}(t)$ has an ultraviolet divergence that can be resolved by an appropriate subtraction \cite{AW2}. However, generally this part is neglible compared to the contribution that depends on $N_{n}(t)$ as the photons are at room temperature. Thus, in the following we neglect this so-called quantum contribution.
\newline
\indent To find the number of excitations $N_{j}(t)$, we solve the following coupled system of equations
\begin{align}
\frac{\partial N_{\mathrm{s}}(t)}{\partial t} &+  \sum_{j} \frac{\partial N_{j}(t)}{\partial t} = 0, \\ \nonumber
\frac{\partial N_{j}(t)}{\partial t} &= N^{2}_{\mathrm{s}}(t) \sum_{j^{\prime}} W_{j,j^{\prime}} (1 + N_{j^{\prime}}(t))(1 + \delta_{j,j^{\prime}} + N_{j}(t)),
\end{align}
where $N_{\mathrm{s}}(t)$ denotes the number of scissors mode quanta and $W_{j,j^{\prime}}$ is the decay rate that is defined in Eq.\,\eqref{eq:probnoalpha}. As a lower limit we neglect the Bose stimulation factors in the rate equations and in this approximation we find 
\begin{align}
N_{j}(t) &= \frac{N^{2}_{\mathrm{s}}(0) \sum_{j^{\prime}} W_{j,j^{\prime}} (1 + \delta_{j,j^{\prime}}) t}{1 + N_{\mathrm{s}}(0) \sum_{j,j^{\prime}} W_{j,j^{\prime}}  (1 + \delta_{j,j^{\prime}}) t}.
\end{align}
Now we use the decay rates as displayed in Fig.\,\ref{fig:Goldenrule} and as an illustration we consider a time $t$ such that $N_{j}(t) = W_{j} / W_{\mathrm{max}}$. The corresponding contribution of the decay of the scissors modes to the density-density correlation function is displayed in Fig.\,\ref{fig:DDcor}. Note that in experiments there is always another contribution from the thermal background, which is given by replacing $\left \langle b_{n}^{\dagger}(t) b_{n}(t) \right\rangle$ by the Bose-Einstein distribution function at energy $\hbar\omega_{n}$. Therefore, to obtain the result of this figure experimentally, the contribution from the thermal background should be subtracted. 
\newline
\indent This is only possible when the contribution originating from this decay process is large enough compared to the background contribution. If we consider a time $t$ such that $N_{j}(t) = W_{j} / W_{\mathrm{max}}$, for $x \simeq x^{\prime}$ the value of the background contribution is at least two orders of magnitude larger and therefore we expect that in this region it is difficult to observe. However, for $|\tilde{x} - \tilde{x}^{\prime}| \geq 0.2$ the largest value of the thermal background is maximally a factor of ten larger than the scissors mode contribution. The exact time scale at which the number of excitations $N_{i}(t)$ takes this value depends on many parameters such as the rotation angle $\chi$, the number of scissors modes and the trapping frequencies. Since in the current experiments the trapping frequencies are of the order of GHz, we obtain that this condition can be fulfilled within the nanosecond regime. 
\newline
\indent We calculated the number of excitations $N_{n}(t)$ while ignoring the Bose stimulation factors in the rate equations. To compare with experimental results, the incorporation of these additional terms can be important. In first approximation these factors can be incorporated by replacing the number of excitations by their expectation value. As this renormalization increases the decay rate, we obtain that within the nanosecond regime the signal of the decay of the scissors modes is comparable to the background for $|\tilde{x} - \tilde{x}^{\prime}| \geq 0.2$. Hence, we expect that the effect of the decay of the scissors modes can be distinguished from the background and is observable in the density-density correlation function.
\newline
\indent In the elongated configuration, as envisaged here, the Bogoliubov amplitudes are proportional to Legendre polynomials. Since these are approximately standing waves, the scissors mode decays predominently into a pair of excitations consisting of an excitation with a certain local momentum $k$ and $-k$. This then also explains the correspondence with the dynamical Casimir effect as an external perturbation, in this case a sudden rotation of the trap, leads to the creation of phonon pairs from the vacuum. Therefore, a measurement of this density-density correlation function would give a demonstration of an analog of the dynamical Casimir effect in a Bose-Einstein condensate of light, which up to now only is considered in atomic and exciton-polariton condensates \cite{DCE1, DCE2, DCE3}.

\section{Discussion and conclusion} 
\label{sec:concl}
In this work we discussed a model for a semiconductor that qualitatively contains the correct crossover physics. However, to obtain the correct quantitative result, several improvements have to be made. First, in our model we only take into account one conduction and valence band. In a realistic semiconductor there are multiple bands that all have to be included. Moreover, we have not taken into account the band-gap renormalization when self-consistently determining the chemical potential. However, for quantitative agreement it will be important to be more careful about this and include the effect of Coulomb screening on the renormalization of the band gap. 
\newline
\indent Another simplification is the use of the contact interaction for the interactions between the electrons and holes. Normally, these interactions are described by a Yukawa potential due to screening effects in the semiconductor. 
Although we use the scattering length for the Yukawa potential as input for the strength of the contact interaction, taking a contact interaction for the interactions between the electrons and holes is still a rough approximation at low carrier densities where screening is not very effective. However, we are primarily interested in large carrier densities and in this regime the Yukawa potential behaves more and more like a contact interaction. We verified that in this regime the susceptibility for Yukawa interactions and the contact interaction agree quite well. For a quantitative agreement at all carrier densities, the Yukawa potential has to be taken into account. Another extension of our model is to consider dynamical screening effects, which become important at very high carrier densities \cite{DynSC}.
\newline
\indent In conclusion, we considered Bose-Einstein condensation of light in nano-fabricated semiconductor microcavities. We modeled the semiconductor as a two-band system consisting of electrons and holes that interact via a contact interaction. To incorporate screening effects, we use the scattering length for the Yukawa potential as input parameter for the strength of the contact interaction. We demonstrated that this model contains a qualitative description of the regime with and without excitons. Moreover, we have shown that if we couple light to the semiconductor, for large carrier densities the finite lifetime effects of the photons can be characterized by a single dimensionless parameter $\alpha$, which is proportional to the slope of the imaginary part of the susceptibility at zero energy. Hereafter, we have proposed to probe the superfluidity of the light in the nano-fabricated semiconductor microcavities via the excitation that the scissors modes. By using Fermi's Golden Rule and a variational ansatz to calculate the Bogoliubov amplitudes, we determined the decay rates of the scissors modes into the non-condensed excitations. Finally, we have demonstrated that the density-density correlation function of the excited light fluid exhibits an analog of the dynamical Casimir effect. 
\newline
\indent This work was supported by the Stichting voor Fundamenteel Onderzoek der Materie (FOM), the European Research Council (ERC) and is part of the D-ITP consortium, a program of the Netherlands Organisation for Scientific Research (NWO) that is funded by the Dutch Ministry of Education, Culture and Science (OCW).

\onecolumngrid
\end{document}